\documentclass[aapm,mph,amsmath,amssymb,reprint]{revtex4-2}
\usepackage{xcolor}
\usepackage{graphicx}
\usepackage{dcolumn}
\usepackage{bm}
\begin{document}
\title{\textbf{Plateau-Rayleigh instability of a viscous film on a soft fiber}}\
\author{Bharti}
\affiliation{Univ. Bordeaux, CNRS, LOMA, UMR 5798, F-33400, Talence, France}\\
\affiliation{Department of Mathematics, Mechanics Division, University of Oslo, Oslo 0316, Norway.}
\author{Andreas Carlson}
\affiliation{Department of Mathematics, Mechanics Division, University of Oslo, Oslo 0316, Norway.}
\author{Tak Shing Chan}
\affiliation{Department of Mathematics, Mechanics Division, University of Oslo, Oslo 0316, Norway.}
\author{Thomas Salez}
\affiliation{Univ. Bordeaux, CNRS, LOMA, UMR 5798, F-33400, Talence, France.}
\email{thomas.salez@cnrs.fr}
\date{\today}
\begin{abstract}
We theoretically study the Plateau-Rayleigh instability of a thin viscous film covering a fiber consisting of a rigid cylindrical core coated with a thin compressible elastic layer. We develop a soft-lubrication model, combining the capillary-driven flow in the viscous film to the elastic deformation of the soft coating, within the Winkler-foundation framework. We perform a linear-stability analysis and derive the dispersion relation. We find that the growth rate is larger when the soft coating is more compliant. As such, softness acts as a destabilising factor. In contrast, increasing the thickness of the soft coating reduces the growth rate, due to the dominating geometrical effect.
\end{abstract}
\maketitle

\section*{Introduction}
The breakup of a Newtonian liquid column into droplets due to an instability driven by surface tension~\cite{Donnelly1997,EV08,Chabert2008} is a commonly observed phenomenon, as illustrated for instance by water falling from a tap~\cite{Zhang1997}.  This phenomenon was first investigated by Plateau~\cite{P43} and Rayleigh\cite{R79a} and hence it is known as the Plateau-Rayleigh instability (PRI). It plays an important role in technologies, such as optical-fiber manufacturing~\cite{Deng2011},  coating~\cite{Dav1999,Primkulov2020,Lee2022} and nanometric crystal growth~\cite{Day2015}. Living creatures also make use of this instability to  achieve particular purposes, as evidenced for example from droplet transport on a spider web~\cite{Zheng2010}, water harvesting by plants~\cite{Chen2018}, or the protein-droplet formation on microtubules~\cite{Setru2021}.  
 
A liquid film coated on a solid cylindrical fiber can also undergo the Plateau-Rayleigh instability~\cite{Goren1962,Goren1964}. Studies have for instance addressed situations where the liquid film is flowing down a vertical fiber~\cite{KLIAKHANDLER2001,Duprat2007,Duprat2009,GONZLEZ2010}. Others have employed the instability to investigate physical effects at interfaces, such as the slip boundary condition~\cite{Haefner2015,Zhang2020a,Zhao2023} at the solid surface, as well as thermal fluctuations~\cite{Zhao2020}, van der Waals forces~\cite{MATTIA2012,Zhao2018,Zhao2019} and surfactant effects at the liquid-air interface~\cite{Carroll1974}, or to trigger droplet motion~\cite{Haefner2015b}. The instability is also present in the late stages of thin-film dewetting~\cite{Brochard1992,Sharma1996, Baumchen2014}, and it can involve viscoelasticity~\cite{Wagner2005,Clasen2006,Deblais2020} as well as non-trivial geometries~\cite{Pairam2009}. Certain studies go beyond the linear-stability analysis and address the non-linear growth dynamics~\cite{Hammond1983, Frenkel1992, Eggers1997}. However, there has been a lack of investigations of the PRI when the solid interface is  deformable. Interestingly, soft solids such as gels and elastomers are ubiquitous in material and biological sciences and their coupling with fluid flows is a current topic of research, as evidenced by soft wetting~\cite{KUMAR2004,Andreotti2020}, soft levelling~\cite{Rivetti2017}, soft dip-coating~\cite{Bertin2022}, contactless rheology of soft materials~\cite{Wang2015,Garcia2016,Karan2018}, or soft-lubrication lift~\cite{Bureau2023,Rallabandi2024}. In this article, we theoretically reveal and characterize the PRI of a viscous film on a fiber consisting of a rigid cylindrical core coated with a thin compressible elastic layer. 
   
\section*{\textbf{Model}}
\begin{figure}[!thbp]
\begin{center}
{\includegraphics[height=1.7 in]{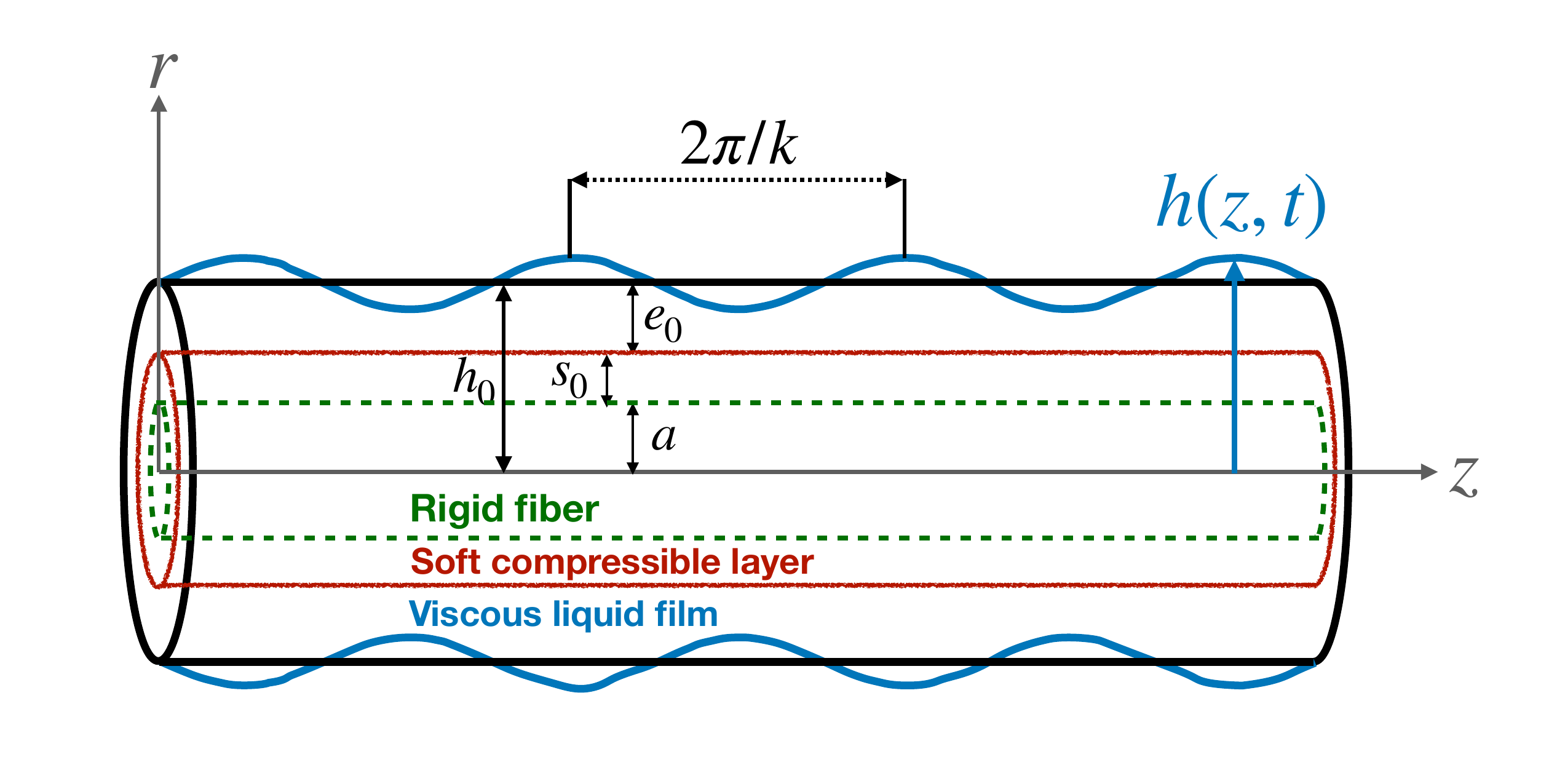}}
\caption{Schematic of the problem. A rigid cylindrical fiber of radius $a$ is pre-coated with a compressible elastic layer of thickness $s_0$, and the ensemble is covered with a thin viscous film of thickness $e_0$. Due to the interplay between capillarity and geometry, the free interface initially located at radial coordinate $r=h_0=a+s_0+e_0$ undergoes a Plateau-Rayleigh Instability (PRI).The latter leads to a viscous flow, which is coupled to the  elastic deformation of the soft coating and gives rise to a radial thickness profile $r=h(z,t)$ along the axial coordinate $z$ and time $t$. A given profile mode is characterized by its angular spatial frequency $k$.}
\label{fig:1}
\end{center}
\end{figure}   
A schematic of the problem is shown in Fig.~\ref{fig:1}. We consider a thin viscous liquid film of initial thickness $e_{0}$, covering an infinitely-long cylindrical rigid fiber of constant radius $a$ coated by a thin, soft and compressible layer of Lam\'e coefficients $G$ and $\lambda$ with initial thickness $s_{0}$. The central axis of the fiber, oriented along the $z$ direction, is chosen as an origin for the radial coordinate $r$. We assume the orthoradial invariance of the free surface profile described by $r = h(z, t) = a + s(z, t) + e(z, t)$, along time $t$. Initially, at $t = 0$, one has $h(z, 0) = h_{0} = a + s_{0} + e_{0}$. 

We now describe the temporal evolution of the system through an incompressible flow of a viscous Newtonian fluid where gravity,  disjoining pressure and inertia are negligible. We consider the local hydrodynamic pressure field $p(r, z, t)$, in excess with respect to the atmospheric one, as well as the local fluid velocity field $\emph{\textbf{v}}(r, z, t)$. We further denote by $\eta$ the shear viscosity and by $\gamma$ the air-liquid surface tension, both taken as homogeneous and constant parameters. In addition, we assume that the lubrication approximation~\cite{Reynolds1886,Oron1997,Craster2009} is valid: the profile slopes remain small at early times (\textit{i.e.} when the linear analysis will be performed), and the velocity is mainly oriented along the fiber axis, \textit{i.e.} $\emph{\textbf{v}} = u(r, z, t) \emph{\textbf{e}}_{z}$, since the horizontal length scale is larger than the minimal wavelength $\lambda_{\textrm{min}}= 2\pi h_{0}$ of the classical PRI that we assume to satisfy $\lambda_{\textrm{min}}\gg e_{0}$.

Let us non-dimensionalize the problem through
$H=h/a$, $H_0=h_0/a$, $S=s/a$, $S_0=s_0/a$, $R=r/a$, $Z=z/a$, $K=ka$, $T=t\gamma/(\eta a)$, $U=u\eta/\gamma$, $P=pa/\gamma$,
and introduce the elastocapillary number
$\alpha=\gamma/[a(2G+\lambda)]$. The flow is well described by Stokes’ equations:
\begin{equation}\label{1}
\partial_{Z}P=\Delta_R U\ ,
\end{equation}
\begin{equation}\label{2}
\partial_{R}P=0\ ,
\end{equation}
where $\Delta_R$ is the radial component of the Laplace operator in cylindrical coordinates. The excess pressure is thus invariant in the radial direction. Thus, according to the Young-Laplace boundary condition at $R=H(Z,T)$, one has:
\begin{equation}\label{3}
P=H^{-1}-H^{''}\ ,
\end{equation}
where the prime denotes from now on the partial derivative with respect to $Z$, and where we recognize the axial and orthoradial curvatures of the free surface in the lubrication approximation. Furthermore, we consider the case of no shear at the liquid-air interface:
\begin{equation}\label{4}
\partial_{R}U|_{R=H}=0\ . 
\end{equation} 
In addition, we assume a no-slip boundary condition at the elastic-liquid interface:
\begin{equation}\label{5}
U|_{R=1+S}=0\ .
\end{equation}

We then integrate Eq.~(\ref{1}), invoking incompressibility and the above-mentioned boundary conditions, and we get:
\begin{equation}\label{6}
U=\frac{H^{'}+H^{2}H^{'''}}{4H^{2}}\left[2H^{2}\log\left(\frac{R}{1+S}\right)-R^{2}+(1+S)^{2}\right].
\end{equation} 
Finally, volume conservation leads to the thin-film equation:
\begin{equation}\label{7}
\dot{H}-\frac{1+S}{H}\dot{S}+\frac{Q^{'}}{H}=0\ ,
\end{equation}
where the dot indicates the partial derivative with respect to $T$, and where we introduced the volume debit per radian:
\begin{eqnarray}
\label{8}
Q&=&\int_{1+S}^{H}\textrm{d}R\, RU\\\label{9}
&=&\frac{H^{'}+H^{2}H^{'''}}{16}\Big[4H^{2}\log\left(\frac{H}{1+S}\right)-3H^{2}\\
&+&4(1+S)^{2}-\frac{(1+S)^{4}}{H^{2}}\Big]\ ,
\end{eqnarray}
which contains the particular no-slip case for a purely rigid fiber~\cite{Haefner2015} when $S = 0$. We stress that the above thin-film equation is in fact a composite equation since we have kept a second-order lubrication term in the
pressure contribution through the axial curvature~\cite{Haefner2015}, in order to counterbalance the driving radial curvature. 

In order to describe the elastohydrodynamic coupling, we assume the deformation-pressure relation at the soft substrate to be linear and local, as in the so-called Winkler foundation~\cite{Dillard2018}. Beyond being a mathematically-convenient way to introduce a minimal elastic coupling, the Winkler foundation is in fact a valid description for thin-enough compressible elastic films~\cite{Chandler2020}. Hence, we write:
\begin{equation}\label{10}
S\propto (1-\alpha P)\ .
\end{equation}
The missing prefactor in the last proportionality relation can be set through the initial condition $S =S_{0}$, and by invoking Eq.~(\ref{3}), so that:
\begin{equation}\label{11}
S=\frac{S_{0}}{1-\frac{\alpha}{H_{0}}}\Big(1-\frac{\alpha}{H}+\alpha H^{''}\Big)\ .
\end{equation}

Let us now perform a linear-stability analysis, by injecting the following mode in Eq.~(\ref{7}): 
\begin{equation}\label{12}
H=H_{0}\left[1+\epsilon\exp(iKZ+\Sigma T)\right]\ ,
\end{equation}
with $\epsilon <<1$ and where $\Sigma$ is the dimensionless modal growth rate. This leads to a second-order low-pass-filter behaviour, with the following softness-induced gain $G$ in the dimensionless dispersion relation:
\begin{equation}\label{13}
G=\frac{\Sigma}{\Sigma_0}=\frac{1}{1-\frac{(1+S_{0})S_{0}\alpha}{H_{0}-\alpha}(H_{0}^{-2}-K^{2})}\ ,
\end{equation}
where we have introduced the reference dimensionless dispersion relation for a rigid fiber of dimensionless radius $1+S_0$:
\begin{equation}\label{14}
\begin{split}
\Sigma_0 &=\frac{(H_{0}^{-2}-K^{2})K^{2}H_{0}}{16}\Big[4H_{0}^{2}\log\left(\frac{H_{0}}{1+S_{0}}\right)-3H_{0}^{2}\\
& +4(1+S_{0})^{2}-\frac{(1+S_{0})^{4}}{H_{0}^{2}}\Big]\ .
\end{split}
\end{equation}

Finally, we identify the fastest-growing mode by differentiating $\Sigma$ with respect to the angular spatial frequency $K$ and equating the result to zero. Doing so, we get the dimensionless angular spatial frequency of the fastest-growing mode, as:
\begin{equation}\label{15}
K_{\textrm{max}}=\sqrt{\frac{1}{H_0^2}-\frac{(H_0-\alpha)\left[1-\sqrt{1-\frac{\alpha S_0(1+S_0)}{H_0^2(H_0-\alpha)}}\,\right]}{\alpha S_0(1+S_0)}}\ ,
\end{equation}
and, subsequently, by injecting the latter in Eqs.~(\ref{13}) and~(\ref{14}), the corresponding dimensionless growth rate $\Sigma_{\textrm{max}}=\Sigma(K_{\textrm{max}})$. Both $K_{\textrm{max}}$ and $\Sigma_{\textrm{max}}$ depend on three parameters: $\alpha$, $S_0$ and $H_0$, thus indicating the fine interplay between geometry and elastocapillarity in this problem.

\section*{Results}
\begin{figure}[!thbp]
\begin{center}
{\includegraphics[height=2.05 in]{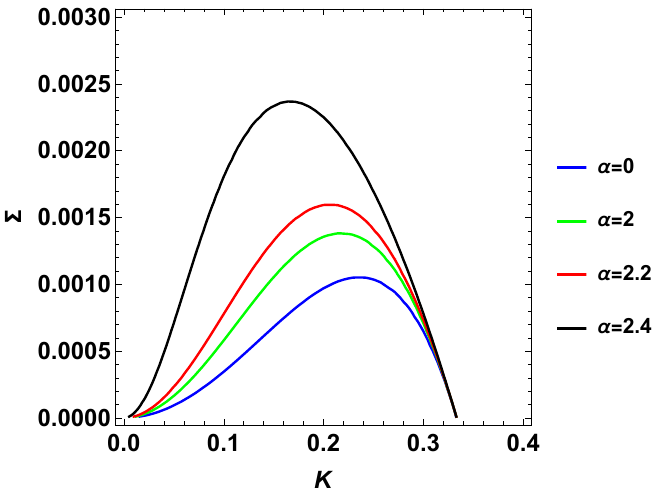}}
\caption{Dimensionless growth rate $\Sigma$ of the PRI (see Eqs.~(\ref{13}) and~(\ref{14})) as a function of the dimensionless angular spatial frequency $K$, for a fixed geometry through $H_{0}=3$ and $S_{0}=1$, and four elastocapillary numbers $\alpha$, as indicated.}
\label{fig:2}
\end{center}
\end{figure}

First, we investigate how the dispersion relation depends on the elastocapillary number $\alpha$. To do so, in Fig.~\ref{fig:2}, we plot from Eqs.~(\ref{13}) and~(\ref{14}) the dimensionless growth rate as a function of the dimensionless angular spatial frequency for different values of $\alpha$, and  fixed values of the geometrical parameters $H_0$ and $S_0$. Naturally, when $\alpha=0$, we recover the result for a purely rigid and no-slip fiber~\cite{Haefner2015}. However, softness appears to amplify the instability rate for any mode $K$. Specifically, with the increase in $\alpha$, the dimensionless growth rate $\Sigma_{\textrm{max}}$ of the fastest-growing mode $K_{\textrm{max}}$ increases. Besides, it is interesting to note that $K_{\textrm{max}}$ decreases with increasing $\alpha$.

\begin{figure}[!thbp]
\begin{center}
{\includegraphics[height=2.05 in]{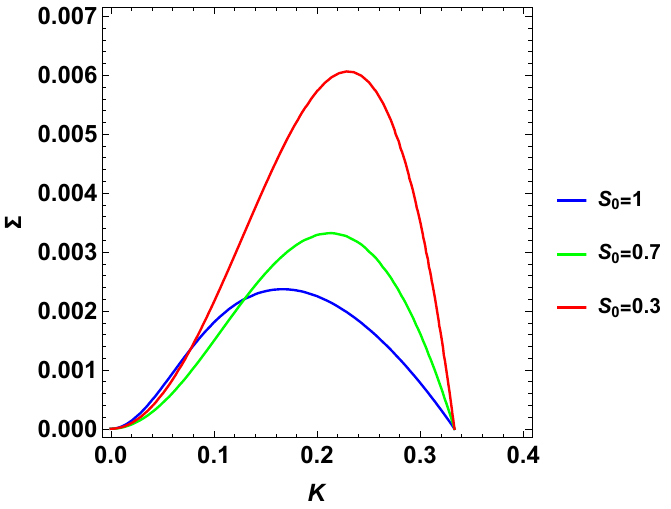}}
\caption{Dimensionless growth rate $\Sigma$ of the PRI (see Eqs.~(\ref{13}) and~(\ref{14})) as a function of the dimensionless angular spatial frequency $K$, for fixed total radius $H_{0}=3$ and elastocapillary number $\alpha=2.4$, and three dimensionless thicknesses $S_0$ of the soft layer, as indicated.}
\label{fig:3}
\end{center}
\end{figure}

\begin{figure}[!thbp]
\begin{center}
{\includegraphics[height=2.05 in]{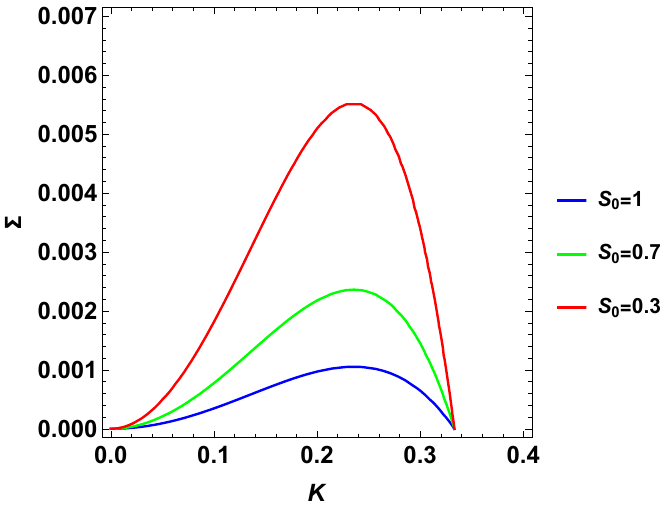}}
\caption{Dimensionless growth rate $\Sigma$ of the PRI (see Eqs.~(\ref{13}) and~(\ref{14})) as a function of the dimensionless angular spatial frequency $K$, for fixed total radius $H_{0}=3$ and elastocapillary number $\alpha=0$, and three dimensionless thicknesses $S_0$ of the soft layer, as indicated.}
\label{fig:4}
\end{center}
\end{figure}

Then, we investigate how the dispersion relation depends on the dimensionless thickness $S_0$ of the soft layer.  In Fig.~\ref{fig:3}, we plot from Eqs.~(\ref{13}) and~(\ref{14}) the dimensionless growth rate as a function of the dimensionless angular spatial frequency for different values of $S_0$, and fixed values of $H_0$ and $\alpha$. Interestingly, we find that the dimensionless growth rate $\Sigma_{\textrm{max}}$ of the fastest-growing mode $K_{\textrm{max}}$ decreases with increasing $S_0$. It may seem contradictory with our above finding that a softer layer destabilises more the liquid film. This can however be understood by the fact that increasing $S_0$ at constant $H_0$ not only increases the compliance of the soft layer but also reduces the liquid-film thickness which slows down the instability~\cite{Haefner2015}. 
This explanation is explicit in Fig.~\ref{fig:4}, where we plot similar results as that of Fig.~\ref{fig:3} but for an elastocapillary number $\alpha=0$, so that the focus is only on geometrical effects.  

\begin{figure}[!thbp]
\begin{center}
{\includegraphics[height=2.05 in]{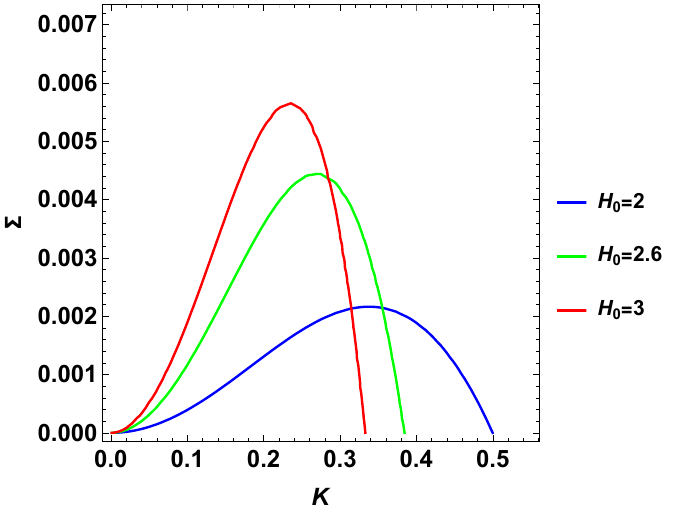}}
\caption{Dimensionless growth rate $\Sigma$ of the PRI (see Eqs.~(\ref{13}) and~(\ref{14})) as a function of the dimensionless angular spatial frequency $K$, for fixed dimensionless thickness $S_0=0.3$ of the soft layer and elastocapillary number $\alpha=1.5$, and three dimensionless total thicknesses $H_0$, as indicated.}
\label{fig:5}
\end{center}
\end{figure}

Next, we investigate how the dispersion relation depends on the dimensionless total thickness $H_0$.  In Fig.~\ref{fig:5}, we plot from Eqs.~(\ref{13}) and~(\ref{14}) the dimensionless growth rate as a function of the dimensionless angular spatial frequency for different values of $H_0$, and fixed values of $S_0$ and $\alpha$. We see that with an increase in $H_0$ the dimensionless growth rate $\Sigma_{\textrm{max}}$ of the fastest-growing mode $K_{\textrm{max}}$ increases. In the mean time, $K_{\textrm{max}}$ decreases with increasing $H_0$. Moreover, the maximal $K$ value for the mode to be unstable (\textit{i.e.} positive $\Sigma$) decreases, as expected from the change of perimeter in the classical PRI.  

\begin{figure}[!thbp]
\begin{center}
{\includegraphics[height=2.05 in]{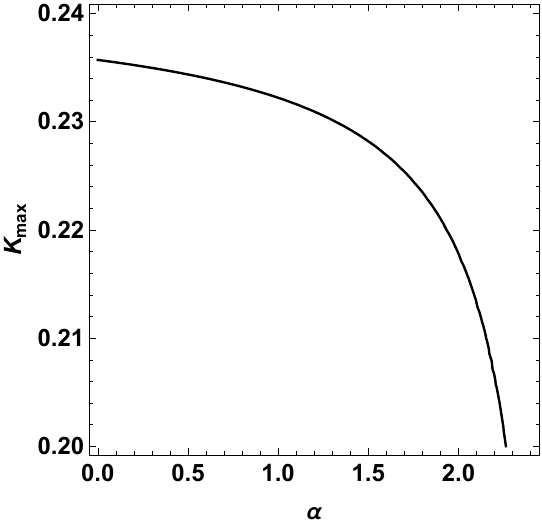}}
\caption{Dimensionless angular spatial frequency $K_{\textrm{max}}$ of the fastest-growing mode of the PRI (see Eq.~(\ref{15})) as a function of the elastocapillary number $\alpha$, for a fixed geometry through $H_{0}=3$ and $S_{0}=1$.}
\label{fig:6}
\end{center}
\end{figure}

\begin{figure}[!thbp]
\begin{center}
{\includegraphics[height=2.05 in]{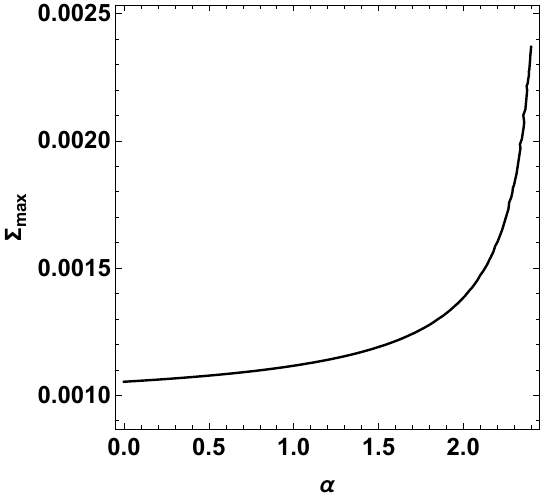}}
\caption{Dimensionless growth rate $\Sigma_{\textrm{max}}=\Sigma(K_{\textrm{max}})$ of the fastest-growing mode of the PRI (see Eqs.~(\ref{13}), (\ref{14}), and~(\ref{15})) as a function of the elastocapillary number $\alpha$, for a fixed geometry through $H_{0}=3$ and $S_{0}=1$.}
\label{fig:7}
\end{center}
\end{figure}

Let us finally focus on the influence of elasticity on the fastest-growing mode, as the latter is the typically-observed one in practice. From Eq.~(\ref{15}), we compute and plot in Fig.~\ref{fig:6} the dimensionless angular spatial frequency of the fastest-growing mode as a function of the elastocapillary number, for a fixed geometry. 
The results confirm, and quantify further, the observation made in Fig.~\ref{fig:2} that the dimensionless typical wavelength of the instability increases with the elastocapillary number. As such, elastic coatings might serve as a way to control the size of droplets produced from the PRI. Moreover, as shown in Fig.~\ref{fig:7}, we also recover that the associated dimensionless growth rate increases with the elastocapillary number. As such, elastic coatings might serve as a way to speed up droplet production from the PRI.

\section*{Conclusion}
We theoretically studied the Plateau-Rayleigh instability of a viscous liquid film on a fiber consisting of a rigid core pre-coated by a compressible elastic layer. By combining lubrication theory and a model Winkler elastic response, we constructed an approximate thin-film equation governing the spatiotemporal evolution of the total radial profile. Linear analysis further provided the modal growth rate of the instability, as a function of the elastocapillary number and the key geometrical aspect ratios of the problem. The principal outcomes are twofold: softness increases i) the typical growth rate, and ii) the typical wavelength of the instability. Our results thus indicate that soft coatings might help designing optimized strategies for patterning and droplet production. While the Winkler foundation is in fact more than a simple toy model, as it actually characterizes well the response of thin compressible elastic layers, it would be interesting to apply similar ideas to other elastic, poroelastic and/or viscoelastic responses, in order to rationalize future experiments and uncover potential novel effects.

\begin{acknowledgments}
The authors thank Oliver B\"aumchen and Virgile Thi\'evenaz for preliminary experimental tests of the model, as well as Elodie Millan for help on the figures. They acknowledge financial support from the European Union through the European Research Council under EMetBrown (ERC-CoG-101039103) grant. Views and opinions expressed are however those of the authors only and do not necessarily reflect those of the European Union or the European Research Council. Neither the European Union nor the granting authority can be held responsible for them. Besides, the authors acknowledge financial support from the Agence Nationale de la Recherche under the Softer (ANR-21-CE06-0029) and Fricolas (ANR-21-CE06-0039) grants, as well as from the Research Council of Norway (project No. 315110). Finally, they thank the Soft Matter Collaborative Research Unit, Frontier Research Center for Advanced Material and Life Science, Faculty of Advanced Life Science at Hokkaido University, Sapporo, Japan. 
\end{acknowledgments}

\bibliography{Bharti2023}
\end{document}